# A Paradigm for Spreadsheet Engineering Methodologies

Thomas A. Grossman
University of San Francisco, School of Business & Management, San Francisco CA 94117-1045
tagrossman@usfca.edu

Özgür Özlük
ISBA Department, College of Business, SFSU San Francisco, CA 94132
ozgur@sfsu.edu

**ABSTRACT**

Spreadsheet engineering methodologies are diverse and sometimes contradictory. It is difficult for spreadsheet developers to identify a spreadsheet engineering methodology that is appropriate for their class of spreadsheet, with its unique combination of goals, type of problem, and available time and resources. There is a lack of well-organized, proven methodologies with known costs and benefits for well-defined spreadsheet classes. It is difficult to compare and critically evaluate methodologies. We present a paradigm for organizing and interpreting spreadsheet engineering recommendations. It systematically addresses the myriad choices made when developing a spreadsheet, and explicitly considers resource constraints and other development parameters. This paradigm provides a framework for evaluation, comparison, and selection of methodologies, and a list of essential elements for developers or codifiers of new methodologies. This paradigm identifies gaps in our knowledge that merit further research.

## 1. INTRODUCTION

Our goal is to see spreadsheet research mature into an important, widely-respected field, which generates research results that are routinely used in business. This goal will be achieved when spreadsheet developers regularly consider which spreadsheet engineering methodology they will apply to a particular spreadsheet. A spreadsheet engineering methodology provides prescriptive recommendations for the choices made throughout the lifecycle of a spreadsheet. Four barriers must be overcome to achieve this goal.

The first barrier is lack of a compelling value proposition. Despite the extensive research on spreadsheet errors [Panko 1998] and the occasional major error that appears in the business press, the argument that current spreadsheet development practices are risky is having little impact. As [Pettifor 2003] points out, "the world is not falling apart through spreadsheet errors". The argument that something must be done about spreadsheet errors is so far achieving little traction in the business world.

With the benefit of hindsight, the limited success of the errors/risks argument is not surprising. In essence, this is an attempt to sell a problem. People don't invest in problems, they invest in solutions. For spreadsheet research to have impact on business practice, it must look beyond errors and their consequences, to the creation of solutions. These solutions—spreadsheet engineering methodologies—must have a compelling value proposition so that busy spreadsheet developers will invest in learning and applying them. An attractive value proposition must include benefits that are important to spreadsheet developers who are relatively unconcerned with risks and errors. These might include more enjoyable development, greater job satisfaction, cost and personnel savings, reduced development time, lifecycle productivity, or enhanced quality of analysis and insight. The contents of this value proposition is an important research question that can only be answered by empirical research on spreadsheet developers.

We note that even if the spreadsheet research community were to be successful in persuading spreadsheet developers and senior management to take spreadsheet risks seriously, those risks can be mitigated only by the use of appropriate spreadsheet engineering methodologies. The barriers below militate against such use.

The second barrier is lack of knowledge of spreadsheet practice. Spreadsheets are undoubtedly the most widely used programming language, and are used for countless different purposes with wide variety in development practices. Unfortunately, there is no systematic knowledge about this diversity of usage and development. This diversity makes it difficult to develop useful generalizations or theories regarding spreadsheets. As discussed by [Grossman and Özlük 2003], any recommendations or theories will apply to only a particular class of spreadsheets with similar characteristics. Empirical research is needed to identify the most important spreadsheet classes so that suitable spreadsheet engineering methodologies can be devised.

The third barrier is lack of a roadmap to appropriate spreadsheet engineering methodologies. Take the point of view of a developer about to embark on a spreadsheet development project. The developer has a certain amount of time and other resources available, is working on a particular type of problem, and has certain (perhaps vaguely defined) goals. What spreadsheet engineering methodology should he adopt?

The current spreadsheet engineering literature is not easily accessible to such a developer. The developer must select among multiple methodologies. It can be difficult to understand which practices are appropriate to a particular spreadsheet, and to match the resources required by a methodology to the resources available. Indeed, existing spreadsheet engineering recommendations are sometimes contradictory, because different spreadsheet classes require different strategies. There is a need for a roadmap that starts with the spreadsheet class and resource constraints and guides developers to appropriate methodologies. We need a theoretical framework, or *paradigm*, to rigorously and systematically organize and critically evaluate spreadsheet engineering methodologies, including identification of their classes, benefits, and resource implications.

The fourth barrier is a lack of well-organized, proven solutions with known costs and benefits for well-defined classes. Developing a roadmap requires a portfolio of spreadsheet engineering methodologies from which developers can choose. To do this, various spreadsheet classes must be identified, and provided with appropriate spreadsheet engineering methodologies. These methodologies must then be compared to alternative methodologies to elucidate when they are most appropriate. Finally, the methodologies need to be tested or otherwise proven to be beneficial, and the proven benefits and demonstrated resource needs must be clearly stated. This is a significant, long-term challenge for spreadsheet researchers. We provide a paradigm of spreadsheet engineering that facilitates the efficient development of spreadsheet engineering methodologies by identifying a set of essential elements that any methodology must consider.

## 2. ESSENTIAL ELEMENTS OF THE PARADIGM

In this section we present a nine-element paradigm for spreadsheet engineering methodologies that facilitates organizing, interpreting, and critically evaluating spreadsheet engineering recommendations. This paradigm provides a vehicle to compare and contrast different spreadsheet engineering recommendations to aid developers in selecting methodologies, and researchers in understanding and improving them. It provides for explicit statements about the relevant classes, and the resources necessary to use the methodology. This paradigm enables us to evaluate the completeness of a spreadsheet engineering methodology. It provides a list of essential elements to developers of new spreadsheet engineering techniques and codifiers of existing practices.

When working with a spreadsheet, the developer makes a series of choices about what to do and how to do it, such as how to organize the cells in the spreadsheet, and what documentation to provide. These choices can be made consciously or unconsciously. If made consciously, they can be made with careful analysis and reflection, or with only momentary consideration. In aggregate these choices determine the efficiency of development, the accuracy of the spreadsheet, and the ability to modify the spreadsheet in the future. These choices are the essence of spreadsheet engineering.

A spreadsheet engineering methodology provides prescriptive recommendations for the choices made throughout the lifecycle of a spreadsheet. By identifying, organizing, and labeling these choices, we can create a paradigm of spreadsheet engineering methodologies. We structure these choices into nine *elements*. Every activity in the lifecycle of a spreadsheet fits into one of these elements. The purpose of the paradigm is to clarify and articulate distinct concepts relating to spreadsheet development and usage. Our intention is that any spreadsheet engineering methodology can be mapped into this paradigm. The nine elements of our paradigm are below.

1. Modeling
2. Development Parameters
3. Design
4. Programming
5. Quality Control
6. Debugging
7. Documentation
8. Usage
9. Modification

A meaningful spreadsheet engineering methodology must consider the problem-solving context in which spreadsheets are created and used. Therefore, the scope of spreadsheet engineering begins with the recognition that a spreadsheet shall be used to address a business problem, and includes all spreadsheet activity through to usage and modification of spreadsheet. Note that we do not consider whether a spreadsheet is the "right" software for the problem; the assumption in this paradigm is that the developer has chosen to build a spreadsheet, and will benefit from guidance on using it well.

Our paradigm is correct if every possible spreadsheet engineering methodology can be mapped onto it in only one way. The choice of elements is somewhat arbitrary. What is important is that the elements be individually distinct and collectively exhaustive. We anticipate this paradigm will be refined as spreadsheet engineering research progresses. Any particular spreadsheet engineering methodology may or may not proceed in the same order that the elements are listed. Many methodologies commingle the elements in the interest of efficiency. (This is desirable, but it makes it difficult to compare and evaluate methodologies.) For example, all spreadsheet engineering recommendations suggest that minimal documentation (element 7) such as row and column labels be done during programming (element 4). When mapping a particular methodology onto our paradigm, it will be necessary to disentangle the various elements to distinguish between the principles embodied in the methodology, and the recommended process of applying those principles. In the subsequent sections, we find it useful to distinguish between a "developer" and a "user"; a *developer* is involved with building the spreadsheet through tasks such as choosing column and row labels and writing cell formulas, whereas a *user* simply enter inputs, and observes and interprets outputs.

## 3. ESSENTIAL ELEMENT 1: MODELING

Modeling is the act of determining what the spreadsheet shall do. Modeling is a component of business problem-solving. The need to solve a problem motivates modeling, which in turn motivates computation. A spreadsheet is a visual computer implementation of a mathematical model. The model embodied in any spreadsheet can be written as a set of algebraic equations which can, in principle, be computed by hand, or coded in a procedural computer language. A spreadsheet model—like any model—takes a set of inputs, and computes a set of outputs. Therefore, we formally define *modeling* as determining the inputs and outputs, and detailing how outputs shall be computed from inputs. Modeling includes considerations of the problem domain discussed in [Grossman 2002]. The best overview of modeling in isolation from programming is chapters 1 – 4 of Powell and Baker 2004.

We intentionally avoid the use of the term "specification" in our definition of modeling. A specification, whose roots are the waterfall lifecycle model of traditional software engineering, describes in great detail the function of a computer program prior to programming. This can be a powerful tool when working with procedural computer languages. In contrast, one of the most powerful capabilities of spreadsheets is their capability to program while modeling. It is apparent that most spreadsheets do not have formal specifications, and it is unlikely that spreadsheet developers will become avid specification writers. Therefore, creating a specification is but one choice that a developer can make, and which will often be declined. A key challenge for spreadsheet engineering researchers is to identify those situations where a specification is indeed essential, cost-effective, or otherwise appropriate.

Because spreadsheets are a powerful vehicle for modeling, modeling is often integrated with spreadsheet design and programming. This can obscure the role of modeling as an independent intellectual activity. The relationship between modeling and programming is an essential aspect of any software engineering methodology. This relationship can range from complete separation to complete integration. Methodologies such as the classic waterfall lifecycle model and the use of Jacksonian Structured Programming [Chadwick et al 1999] recommend the completion of modeling before the start of programming. Certain lifecycle models such as the spiral model [McConnell 1996] provide for a sequence of distinct modeling and programming steps. [Nardi and Miller 1991] describe how spreadsheet users and developers cooperate in creating spreadsheets, with programming and modeling partially integrated. [Grossman 2002] discusses how developers can engage in exploratory modeling, where they program a spreadsheet to help them think through and understand their business problem fully integrating modeling and programming.

It is important that any spreadsheet engineering methodology address modeling, which is the process of figuring out what the spreadsheet is to do, and carefully discuss the interaction of modeling and programming.

## 4. ESSENTIAL ELEMENT 2: DEVELOPMENT PARAMETERS

*Development parameters* are the planning assumptions of a spreadsheet. This includes the goals of the spreadsheet; the budget in terms of money, time and developer labor; the users in terms of their number, skill, and experience; the frequency of use; the time period of use; the likelihood and nature of modifications after usage; interactions with other information systems; the importance of the spreadsheet; the desired accuracy; and any other considerations that may affect the spreadsheet during its lifecycle.

The selection of development parameters strongly affects all the steps of spreadsheet development. Unfortunately, because development parameters are prospective, they can be wrong. For example, a spreadsheet intended for one-time usage by its developer might see usage by multiple users. Or a spreadsheet that was to be programmed once and deleted is modified for other uses. Poor selection of development parameters at the beginning can cause expense and risk later. Therefore, the establishment of development parameters includes any evaluation of risks, such as errors and development failure.

Development parameters are essentially business judgments about the deployment of resources to create information systems to achieve organizational goals. Therefore, development parameters are controlled by business considerations and resource constraints, not by any inherent properties of the model to be developed.

We believe that consideration of development parameters is an essential component of any spreadsheet engineering methodology. A given spreadsheet engineering methodology is more appropriate for some development parameters than others. However, there is a tendency in the spreadsheet engineering literature to provide insufficient discussion of development parameters. Spreadsheet engineering methodologies should clearly identify any assumptions of development and usage, and discuss the resources required during initial development and potential future modifications.

## 5. ESSENTIAL ELEMENT 3: DESIGN

The *design* of a spreadsheet comprises two elements: structural design and visual design. *Structural design* is the way cells are arranged. Structural design includes the designation of rows and columns to have particular meaning, the use of modularity, and the provision of space for documentation. *Visual design* refers to the appearance of cells and cell borders. Visual design includes shading, borders, fonts and other formats.

The spreadsheet engineering literature is in agreement that good design is important. However, there is no agreed list of what constitutes good design. Many discussions in the literature mingle design considerations with programming and documentation.

Two principles of structural design are widespread. The first is to organize related concepts using the rows and columns of the spreadsheet. For example, each column of a cash flow statement contains a single year, and each row contains a single accounting concept.

The second structural design principle is "modularity", which says that logically related elements be grouped into modules. A module might contain model inputs, model outputs, a summary with selected inputs and outputs, a set of computations, or other items. The module(s) that a spreadsheet user interacts with are called the "user interface" and often require special attention. A module can comprise a single cell, a section of a worksheet, an entire worksheet, a workbook, or even a set of linked workbooks. Modules can contain submodules. For example, the authors recently observed a 27 MB workbook of a user interface for a large spreadsheet application. The workbook contains numerous submodules in the form of worksheets, with each worksheet containing a number of submodules in the form of sections.

The final structural design principle is to provide space for documentation. Space should be provided for row and column labels, and any other documentation. This is discussed in more detail in the Documentation section below.

Visual design is essentially the formats applied to the spreadsheet. The most important visual design elements are the use of fonts (including font choice, color, and bold/italic), justification within a cell, cell colors, and cell borders. Visual design draws attention to important elements in the spreadsheet; differentiates among inputs, intermediate variables and outputs; communicates and reinforces modularity; and establishes hierarchical relationships within a module.

Visual design elements are important for several reasons. [Nardi and Miller 1990] argue that spreadsheets success relies on the strong visual format opportunities for structuring and presenting data. [Reithel et al 1996] tells us that well-formatted spreadsheets are perceived as more accurate. However, [Raffensperger 2000] argues that certain formats such as excessive color, and non-constant column widths may reduce comprehension or become a distraction.

## 6. ESSENTIAL ELEMENT 4: PROGRAMMING

*Programming* is the creation of cell formulas and other logic in a spreadsheet. Programming techniques range from broad principles such as "build a small-scale prototype and then scale up" to precise recommendations such as "do not hard code constants in cell formulas".

Here is a limited list of programming techniques. Enter each input exactly once. Replicate cell formulas using copying rather than typing. Use absolute references in cell formulas to facilitate copying. Create a scenario tool to run multiple datasets through the same logic. Check that input values are within established ranges. Use data validation tools to prevent users from entering out-of-range inputs. Use cell protection to prevent accidental or unauthorized modification of cell formulas. Check intermediate calculations for out-of-range values. Use cross-foots and other redundant calculations. Use version control. Make files read-only to prevent accidental overwriting of important information. Use spreadsheet productivity features (such as Insert\Function…\, and selecting cell references rather than typing them) to avoid syntax errors in formulas. There are many other techniques.

There are some contradictions and open questions in the literature. Some authorities recommend avoidance of certain spreadsheet functions deemed risky, such as OFFSET, but others recommend OFFSET as being useful. Many spreadsheet auditing packages flag the use of multiply-nested IF functions, yet some well-engineered spreadsheets use deep nesting. The use of range names is recommended, but there are times when range names interfere with copying formulas or modifying a spreadsheet. [Thommes 1994] and [Caine and Robson 1993] recommend splitting lengthy cell formulas into smaller parts to keep the formulas simple and easy to understand, whereas [Raffensperger 2000] argues that splitting formulas may result in bloated hard-to-read spreadsheets.

Spreadsheet programming currently resemble a bag of tricks rather than a well-organized, intellectually coherent toolbox. Research is needed to codify and organize the techniques. Contradictory recommendations should be identified, and these contradictions resolved by specifying the development parameters or design where each technique is appropriate. It would be helpful to bring intellectual coherence to the dizzying array of techniques, by categorizing them and distinguishing among high-level principles and low-level practices. Of particular interest for spreadsheet engineering is how development parameters affect programming practices and how programming practices affect spreadsheet quality and usage.

## 7. ESSENTIAL ELEMENT 5: QUALITY CONTROL

*Quality Control* is all actions taken to determine whether the outputs of a spreadsheet are satisfactory. There are two aspects to quality control, "verification" and "validation". Verification is concerned with the programming of the model, and validation is concerned with the meaningfulness of the model as implemented.

*Verification* evaluates accuracy. It asks whether the spreadsheet program correctly implements the model. Verification answers the question "does the spreadsheet contain programming errors?" In principle, verification is an objective evaluation.

*Validation* evaluates model quality. It asks whether the model adequately depicts reality, and how closely model outputs correspond to real world values. Validation answers the question "is the model adequate?" In some situations, such as simple taxation models, validation can be objectively evaluated. In other situations, such as a complex model used by a bank to set the residual prices on car leases, objective evaluation is impossible.

There are two general approaches to verification, code inspection and testing. The software engineering literature and [Panko 1999] argue that multi-person code inspection has the highest error-detection rate. Spreadsheet auditing tools automate code inspection for certain errors.

Testing is entering test data into a model and observing the outputs. The most useful test inputs are those with corresponding outputs that are known to be correct. For spreadsheets that address a problem that has never previously been modeled, the generation of test cases is a significant challenge. The theoretically rigorous approach to testing in [Rothermel et al 2001] assumes that test cases are available. Probably the best reference on testing spreadsheets that lack test cases is the brief and non-comprehensive discussion in chapter 5 of [Powell and Baker 2004].

In some cases, working through the model logic manually may be the only practical way to obtain a test case. In extreme situations, it may be necessary to build independently a parallel system, and compare results. Future research should consider how to test spreadsheets where the correct outputs are not known *a priori*.

Validation and verification are distinct concepts. An inadequate model that is programmed well is verified and invalid. A satisfactory model that has programming errors is inverified and valid. (Note: a model that has not been evaluated for accuracy is "unverified", a model that has failed the evaluation is "inverified".) Unfortunately, verification and validation are sometimes conflated in practice. When a developer evaluates a spreadsheet by examining the outputs and judging that they "seem about right" they are engaging in validation rather than verification. They risk making the (often unstated) assumption that validation insures verification. It does not. It simply insures that any errors tend towards what the developer expected for the inputs being used. This is called "confirmation bias" and can mask errors in a spreadsheet. There is no research on the prevalence of this practice, but anecdotal evidence suggests it is widespread. In fact, [Burnett et al 1999] take this approach, where testing is performed by developers "noticing" whether a particular cell contains a correct or incorrect value.

When verification detects an error in the programming, or validation detects an error in the model, it is necessary to fix the problem. This is called debugging.

## 8. ESSENTIAL ELEMENT 6: DEBUGGING

Quality control finds problems, and debugging fixes them. *Debugging* is modifying a spreadsheet program to fix an output that has an unsatisfactory value. There are three key issues in debugging: how to locate the source of the problem, how to fix the problem, and how to avoid introducing new bugs.

Unfortunately, the spreadsheet engineering literature contains little guidance on debugging, particularly in a spreadsheet with complex contingent logic programmed with lookup formulas or nested IF formulas. Spreadsheet auditing software can be helpful in locating the problem, and can sometimes provide guidance in fixing it.

When the issue with the spreadsheet is the invalidity of the underlying model, it may be necessary to revisit element 1, modeling, to enhance the model. Then, it is necessary to change the spreadsheet to incorporate the enhancements. Making these changes accurately and efficiently is a similar if not more demanding skill compared to making changes to eliminate a programming error.

The software engineering and quality control literatures argue persuasively that preventing errors is cheaper than finding and fixing errors. Thus, incremental investment in design and programming can bring disproportionate savings in debugging. Provided of course, quality control is done at all!

Development of techniques for debugging is an area that has not received enough attention and merits further research. Empirical research on quality control and debugging practices would be valuable.

## 9. ESSENTIAL ELEMENT 7: DOCUMENTATION

*Documentation* is any written record regarding the spreadsheet. The most common form of documentation is row and column labels in a spreadsheet. These can range from minimal abbreviations, to lengthy formal names.

Documentation can reside in many places. It can be integrated with cell formulas within a module, for example row and column labels, and a notes column. Documentation can reside in its own documentation module within a spreadsheet, or it can be in the form of a separate document such as a full-fledged user's manual. Microsoft Excel has features such as Comments and text boxes that allow documentation to be placed almost anywhere in the spreadsheet. The programming technique of range names can make cell formulas more readable and serves as a form of documentation.

Any element of a methodology can be documented, including modeling, development parameters, design, programming, quality control, and debugging. Documentation can consume substantial resources, and the amount of documentation to be done depends on the development parameters.

Spreadsheet engineering methodologies must carefully consider the appropriate level and type of documentation, and the resources required to create it. It is well known from software engineering that documentation is often inadequate, and there is anecdotal and empirical evidence that spreadsheet documentation is inadequate. Therefore, it may be desirable for methodologies to distinguish between essential documentation and desirable documentation.

## 10. ESSENTIAL ELEMENT 8: USAGE

We define the *usage* of a spreadsheet to be any process where a user provides inputs to a spreadsheet, and observes the outputs. Usage does not involve programming.

Planned usage is considered in the development parameters element. Actual usage may differ from planned usage. This can signal unanticipated success. Such success can be a mixed blessing, because unplanned usage implies the spreadsheet is a poor platform for those newly-discovered uses, whatever they may be.

There is great diversity in usage, but we have limited theoretical and empirical knowledge because spreadsheet usage receives little attention in the literature. Usage can be by the developer, or by other individuals. Usage can be by one individual or many. Usage can be a single observation of a set of model outputs, or can involve multiple sets of inputs used to generate multiple sets of outputs. Usage can be a one-off event, or can take place regularly or irregularly over time. The user may or may not be able to interact with the developer. There are questions about how users can analyze a spreadsheet model (or any model) to systematically extract insight about a business process. There are also questions on how spreadsheets are shared by different people.

This situation is problematic. The usage of a spreadsheet is particularly important to spreadsheet development because expected usage helps determine the development parameters. We are ignorant of usage, and therefore cannot present compelling, evidence-based suggestions for integrating usage expectations into spreadsheet development. We know that usage expectations are sometimes wrong, but we do not know how often or how expected usage correlates to actual usage.

Clearly, rigorous research on spreadsheet usage would be beneficial.

## 11. ESSENTIAL ELEMENT 9: MODIFICATION

*Modification* refers to changes made to the spreadsheet after it has been used. This includes terms such as "maintenance", "enhancement" and "extension". Like any software, modification of spreadsheets can be substantially more expensive than building in features from the beginning, and provision for modifications made early in development can significantly reduce the time, cost and risk of making modifications.

We know little about modifications. We know that spreadsheets whose development parameters indicated no modifications may indeed be modified after usage. Even when modifications are included in the development parameters, it may not be possible to anticipate the nature of the modifications. There is no systematic research on the origin of modification requests, and how these connect to usage, development parameters, design and programming decisions.

There is a clear opportunity for research on spreadsheet modifications. It would be helpful to have a categorization of the kinds of modifications that are made, who proposes them, and the effect of early-development planning (or its lack) on later modifications. In particular, it would be valuable to better understand the twin risks of over-engineering for modifications that never happen vis-à-vis the risk of under-engineering for unexpected modifications that later prove necessary.

## 12. CLASSES OF SPREADSHEETS

Because of the great diversity among spreadsheets and spreadsheet developers, it is difficult to make detailed spreadsheet engineering recommendations that are widely applicable. In contrast, recommendations with narrow scope, pertaining to specific classes of spreadsheets, can provide detailed and specific guidance. The level of detail and specificity of spreadsheet engineering recommendations is inversely proportional to the scope of the recommendations. Therefore, it is important that any spreadsheet engineering methodology carefully define the class of problems to which it applies. In our nine-element paradigm, the defining characteristics of a class are to be found in the development parameters and modeling elements.

When the class is well-defined, highly specific methodologies can be provided. For example, [Conway and Ragsdale 1997] consider the narrowly-defined class of small scale optimization models, and are able to provide very specific recommendations.

Therefore, it is essential that a spreadsheet engineering methodology clearly indicate in the development parameters and modeling elements the class of spreadsheet to which it applies. We note that many existing spreadsheet engineering methodologies provide insufficient class information, and have hidden assumptions about where they can usefully be employed. This reduces their effectiveness, because some of their recommendations are ineffective or even inappropriate in certain classes, and developers for whom the recommendations are most appropriate may not recognize the relevance of the methodology.

## 13. CONCLUSIONS

Our hope is to see spreadsheet research mature into an important, widely-respected field, which generates research results that are extensively used in business. This entails prescriptive research with sufficient power and applicability to motivate adoption and employment by busy spreadsheet developers. This power will emerge only with specific, detailed methodologies. Applicability will emerge with carefully specification of class. In principle, a specific spreadsheet engineering methodology can be defined for any class of spreadsheet. In practice, methodologies will probably only be defined for classes where significant value can be obtained through better practices.

Currently, it is difficult to compare, contrast, and critically evaluate spreadsheet engineering methodologies. This is because the methodologies are organized to support a particular development process, and it is challenging to decompose them into their components to observe commonalities and differences. More important, it can be difficult to recognize the hidden assumptions that underlie many methodologies, particularly assumptions about class.

We present paradigm of spreadsheet engineering that will help with these difficulties. By mapping spreadsheet engineering recommendations onto the nine essential elements of this paradigm, it will be easy to compare and evaluate methodologies, and determine their completeness. This paradigm provides a framework for developing new spreadsheet engineering methodologies, and makes explicit provision for identifying the modeling approach and development parameters which together define the class.

This paradigm identifies gaps in our knowledge that can guide further research. Important research opportunities include systematizing and organizing the wealth of programming techniques; devising techniques for testing spreadsheets that lack test cases; systematic methods for debugging spreadsheets; better understanding spreadsheet usage; and increasing our knowledge of spreadsheet modifications. Finally, there is an opportunity to systematically interpret existing spreadsheet engineering methodologies in light of our nine-element paradigm, and compare models to identify commonalities.

Future research on spreadsheet engineering should identify high-value classes with large numbers of developers who are sensitive to the investment they make in spreadsheets. These are the audiences most likely to adopt new spreadsheet engineering methodologies. Spreadsheet developers should be interviewed to identify what value proposition would induce them to invest in deploying new methodologies in their organizations.

With the class carefully defined, and a clear sense of the benefits that developers need to see, researchers can use our nine-element paradigm to devise an appropriate, detailed spreadsheet engineering methodology, which they can then test against current practice to determine the benefits and required resources.

**REFERENCES**

Burnett, M., Sheretov, A., and Rothermel, G. (1999), "Scaling up a 'What You See is What You Test' Methodology to Spreadsheet Grids", Proceedings of the 1999 IEEE Symposium on Visual Languages, pp. 30-37.

Caine, D. J. and Robson, A. J. (1993), "Spreadsheet modelling: Guidelines for model development", Management Decision 31(1), pp. 38-44.

Chadwick, D., Rajalingham, K., Knight, B., and Edwards, D. (1999), "A Methodology for Spreadsheet Development Based on Data Structure", University of Greenwich Centre for Numerical Modelling and Process Analysis 99(50), pp. 1-12.

Conway, D. G. and Ragsdale, C. T. (1997), "Modeling optimization problems in the unstructured world of spreadsheets", Omega 25(3), pp. 313-322.

Grossman, T. A. (2002), "Spreadsheet Engineering: A Research Framework", European Spreadsheet Risks Interest Group 3rd Annual Symposium, Cardiff, pp. 21-34, July.

Grossman, T. A. and Özlük, Ö. (2003), "Research Strategy and Scoping Survey on Spreadsheet Practices", European Spreadsheet Risks Interest Group 4th Annual Symposium, Dublin, pp. 23-32, July.

McConnell, S. (1996), Rapid Development: Taming Wild Software Schedules, Microsoft Press.

Nardi, B. A. and Miller, J. R. (1990), "The Spreadsheet Interface: A Basis for End User Programming", IFIP TC 13 Third International Conference on Human-Computer Interaction, Cambridge, U.K. Elsevier Science Publishers.

Nardi, B. A. and Miller, J. R. (1991), "Twinkling lights and nested loops: distributed problem solving and spreadsheet development. IJM-Machine Studies 34, pp. 161-184.

Panko, R. R. (1999), "Applying code inspection to spreadsheet testing", Journal of Management Information Systems 16(2), pp. 159-176.

Panko, R. R. (1998), "What We Know About Spreadsheet Errors", Journal of End User Computing, 10(2), pp. 15-21.

Pettifor, B. (2003), "Management Summary: Getting spreadsheets under control—practical issues and ideas", European Spreadsheet Risks Interest Group 4th Annual Symposium, Dublin, pp. 105-110, July.

Powell, S. P. and Baker, K. R. (2004), "The art of modeling with spreadsheets: Management science, spreadsheet engineering and modeling craft", Wiley.

Raffensperger, J. F. (2000), "The new guidelines for writing spreadsheets", http://www.mang.canterbury.ac.nz/people/jfraffen/spreadsheets/index.html, accessed April 9, 2004.

Rothermel, G., Burnett, M., Li, L., DuPuis, C., and Sheretov, A. (2001), "A Methodology for Testing Spreadsheets", ACM Transactions on Software Engineering and Methodology, 10(2).

Thommes, M. C (1994), "Advanced Spreadsheet Design Using Lotus Macros", Boyd & Fraser Publishing Company.